\documentclass[aip,
 amsmath,amssymb,
 preprint,%
]{revtex4-2}

\usepackage{graphicx}
\usepackage{dcolumn}
\usepackage{bm}
\usepackage{ulem} 
\usepackage{colordvi,color}
\usepackage[dvipsnames]{xcolor}  
\usepackage{float}
\usepackage{appendix}
\usepackage{soul,xcolor}  
\usepackage{xspace}
\usepackage{textcase} 

\usepackage[unicode=true,pdfusetitle,
bookmarks=true,bookmarksnumbered=false,bookmarksopen=false,
breaklinks=false,pdfborder={0 0 1},backref=false,colorlinks=false]
{hyperref}

\begin{document}

\renewcommand{\baselinestretch}{1.2}

\renewcommand{\thetable}{\arabic{table}}
\newcommand{\rojo}[1]{{\color{red}{#1}}}
\newcommand{\azul}[1]{{\color{blue}{#1}}}
\newcommand{\magenta}[1]{{\color{magenta}{#1}}}

\newcommand{\isAJ}{i$\sigma$AuxJ\xspace}
\newcommand{\isAMP}{i$\sigma$AuxMP2\xspace}
\newcommand{\isDZ}{i$\sigma$DZ1\xspace}
\newcommand{\isTZ}{i$\sigma$TZ1\xspace}
\newcommand{\isQZ}{i$\sigma$QZ1\xspace}
\newcommand{\isXZ}{i$\sigma$XZ1\xspace}

\newcommand{\rne}{{\bf r}}
\newcommand{\Rne}{\mathbf{R}}

\newcommand{\baa}{\begin{eqnarray}}
\newcommand{\eaa}{\end{eqnarray}}

\newcommand{\be}{\begin{equation}}
\newcommand{\ee}{\end{equation}}
\newcommand{\ba}{\begin{array}}
\newcommand{\ea}{\end{array}}

\title{Serially Improved GTOs for Molecular Applications (SIGMA): basis sets from monovalent ions}

\author{Ignacio Ema L\'opez}
\email{nacho.ema@uam.es}
\affiliation{Departamento de Qu\'imica F\'isica Aplicada. Universidad Aut\'onoma de Madrid. E-28049. Madrid. Spain}

\author{Guillermo Ram\'irez Moreno}
\affiliation{Departamento de Qu\'imica F\'isica Aplicada. Universidad Aut\'onoma de Madrid. E-28049. Madrid. Spain}

\author{Rafael L\'opez Fern\'andez}
\affiliation{Departamento de Qu\'imica F\'isica Aplicada. Universidad Aut\'onoma de Madrid. E-28049. Madrid. Spain}

\author{Jos\'e Manuel Garc\'ia de la Vega}
\affiliation{Departamento de Qu\'imica F\'isica Aplicada. Universidad Aut\'onoma de Madrid. E-28049. Madrid. Spain}

\begin{abstract}
    A new family of ionic basis sets, denoted i$\sigma$XZ1, is presented for molecular calculations on systems containing monovalent ions. The basis sets extend the SIGMA family by explicitly accounting for the different electronic structure of cations and anions. Auxiliary basis sets for resolution-of-the-identity calculations are also developed for the Coulomb term. The performance of the proposed basis sets is assessed for alkali-halide clusters. Compared with conventional basis sets, i$\sigma$XZ1 provides an accurate description of structural, energetic, and electronic properties while showing remarkable robustness against near-linear dependencies, allowing stable calculations on systems containing up to 792 NaCl units. These features make the i$\sigma$XZ1 family a reliable and efficient alternative for large-scale calculations on ionic systems.
\end{abstract}

\maketitle

\section*{Introduction}

A new ionic basis-set family, \isXZ, designed for accurate molecular calculations in ionic systems is presented for the monovalent ions H$^-$, Li$^+$, F$^-$, Na$^+$, Cl$^-$, and K$^+$. This family extends the Serially Improved Gaussian Orbitals for Molecular Applications (SIGMA)\cite{Ema2023,Ema2025a,Ema2025b} basis sets by explicitly accounting for the distinctive features of ionic environments. Basis sets of different cardinalities are reported, together with auxiliary basis sets for the resolution-of-the-identity (RI) approximation of the Coulomb term, denoted \isAJ, whose composition and performance are described in the Appendix. The \isXZ  and \isAJ basis sets in ORCA format are collected in the files {\it iSigmaXZ1.txt} and {\it iSigmaAuxJ.txt} in the Supplementary Information.

The composition of the \isXZ basis sets is summarized in Table~\ref{tab1}. The same table also compares them with the corresponding Dunning basis sets (caXZ), namely the Augmented Correlation-Consistent Polarized Valence X-Zeta (aXZ) basis sets for the anions,\cite{Kendall1992,Woon1993a} the Correlation-Consistent Polarized Core-Valence X-Zeta (cXZ) basis sets for the cations,\cite{Prascher2011} and, for potassium, the Correlation-Consistent Polarized Weighted Core-Valence X-Zeta (pwcXZ) basis sets.\cite{Hill2017} The suffix "1" in the \isXZ basis-set names denotes the additional primitive introduced into each polarization-shell contraction. 

Ground-state energies computed with the \isXZ basis sets are compared with those obtained using the corresponding Dunning basis sets in Table~\ref{tab2}, where the numbers of contracted functions are given in parentheses. As can be seen, despite containing fewer contracted functions, the \isXZ basis sets yield lower CISD energies than the corresponding Dunning basis sets. At the HF level, only a few exceptions are found, for which the Dunning basis sets produce slightly lower energies.

To assess the performance of the \isXZ basis sets in extended ionic systems, (NaCl)$_n$ clusters were investigated using the HF, PBE, and MP2 methods. The results reported in Table~\ref{tab3} show that the PBE method provides an accurate description of (NaCl)$_n$ clusters. Since electron-correlation effects are found to be small for these highly ionic systems, this conclusion is expected to remain valid for the other (MX)$_n$ clusters, where M = Li, Na, or K; X = H, F, or Cl, considered in this work. Consequently, Tables~\ref{tab4}--\ref{tab7} report only RI-PBE results.

It is worth noting that the compactness of the \isXZ basis sets and the carefully chosen exponents of their diffuse primitives make them highly robust against near-linear dependencies, thereby avoiding numerical instabilities\cite{Lehtola2019} even in calculations on very large systems. In particular, the single-point PBE calculation for the largest cluster considered, (NaCl)$_{792}$, involving 96,224 auxiliary and 50,688 basis functions with the \isAJ/\isTZ basis-set combination, converged without any numerical difficulties.

The combination of the \isXZ basis sets with the Deformed Atoms in Molecules (DAM) partition/expansion of the molecular electron density\cite{FernandezRico2004b} offers an additional advantage: the rapid convergence of the molecular electrostatic potential (MESP) when expressed as a sum of atomic contributions.

The electrostatic potential generated by a molecule or ion with $N$
nuclei at point $\rne$ is given, in atomic units, by:

\be   \label{eq:1.1}
V(\mathbf{r}) = \sum_{I=1}^N \frac{\zeta_I}{r_I}
- \int d\mathbf{r}^\prime
\frac{\rho(\mathbf{r}^\prime)}{|\mathbf{r}-\mathbf{r}^\prime|}
\ee
where $\rho(\rne)$ is the one-electron density, $r_I = |\rne-\Rne_I|$,
and $\zeta_I$ is the charge of the nucleus located at
$\Rne_I = (X_I,Y_I,Z_I)$.

According to the DAM method,\cite{FernandezRico2004a} the electron density is expanded as:

\be \label{eq:1.2}
\rho(\mathbf{r}) =
\sum_{I=1}^N \rho^I(\mathbf{r_I})
= \sum_{I=1}^N \sum_{l=0}^\infty \sum_{m=-l}^l
z_l^m(\rne_I) \; f_{lm}^I(r_I)
\ee
where $z_l^m(\rne_I)$ are real regular harmonics:

\be   \label{eq:1.3}
\ba{lr}
z_l^m(\mathbf{r}) = r^l \; (-1)^m \; P_l^{m}(\cos \theta)
\; \cos(m \phi) & \;\;\;\;\;  \mbox{for } \;\;\;\;\; m \ge 0 \\
z_l^m(\mathbf{r}) = r^l \; (-1)^{|m|} \; P_l^{|m|}(\cos
\theta) \; \sin(|m| \phi) & \;\;\;\;\;  \mbox{for } \;\;\;\;\; m < 0
\ea
\ee
where $P_l^{|m|}(\cos \theta)$ are the associated Legendre functions
(see Ref.~\cite{Gradshteyn} eq.~8.751.1), and $f_{lm}(r)$ are radial factors. 

This expansion of the density directly yields the expansion of the MESP:\cite{FernandezRico2004b}

\be \label{eq:1.4}
V(\rne) = \sum_{I=1}^N \left\{ \frac{\zeta_I}{r_I}
- \sum_{l=0}^\infty \sum_{m=-l}^l z_l^m(\rne_I) \; \left[
\frac{Q^I_{lm}(r_I)}{r_I^{2l+1}} + q^I_{lm}(r_I) \right] \right\}
\ee

where, for {\it each atom} (omitting the $I$),

\be   \label{eq:1.5}
Q_{lm}(r)  =  \frac{4 \; \pi}{2l+1} \;
\int_0^{r} dr \; {r}^{2l+2} \; f_{lm}(r)
\ee

and

\be   \label{eq:1.6}
q_{lm}(r) =  \frac{4 \; \pi}{2l+1} \;
\int_{r}^\infty dr \; r \; f_{lm}(r)
\ee
where $Q_{lm}(r)$ and $q_{lm}(r)$ can be efficiently computed as described in Ref.\cite{FernandezRico2004b}.

\begin{figure}[H]
    \hspace*{-1.7cm}\includegraphics[width=0.67\textwidth]{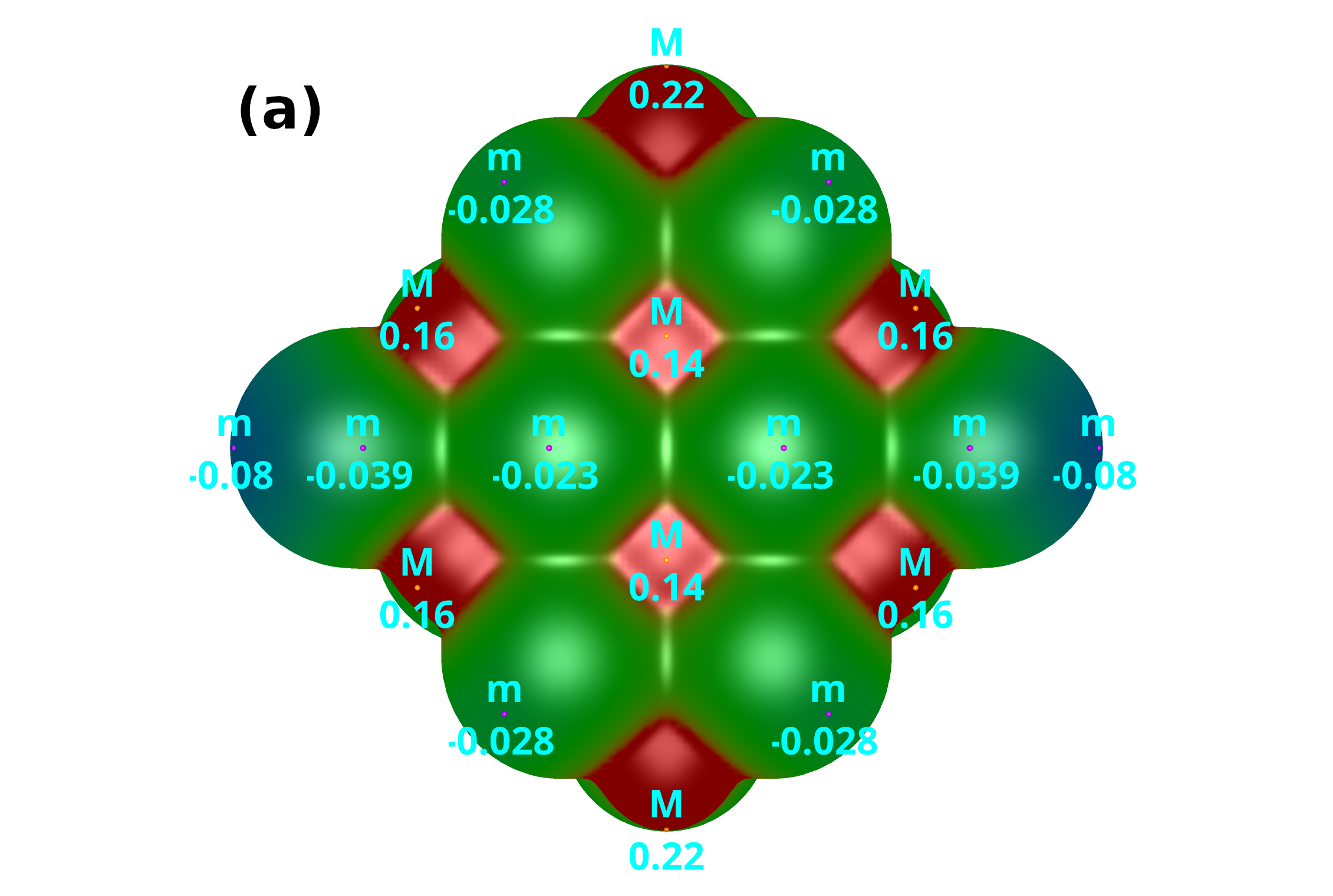}
    \hspace*{-1.5cm}\includegraphics[width=0.67\textwidth]{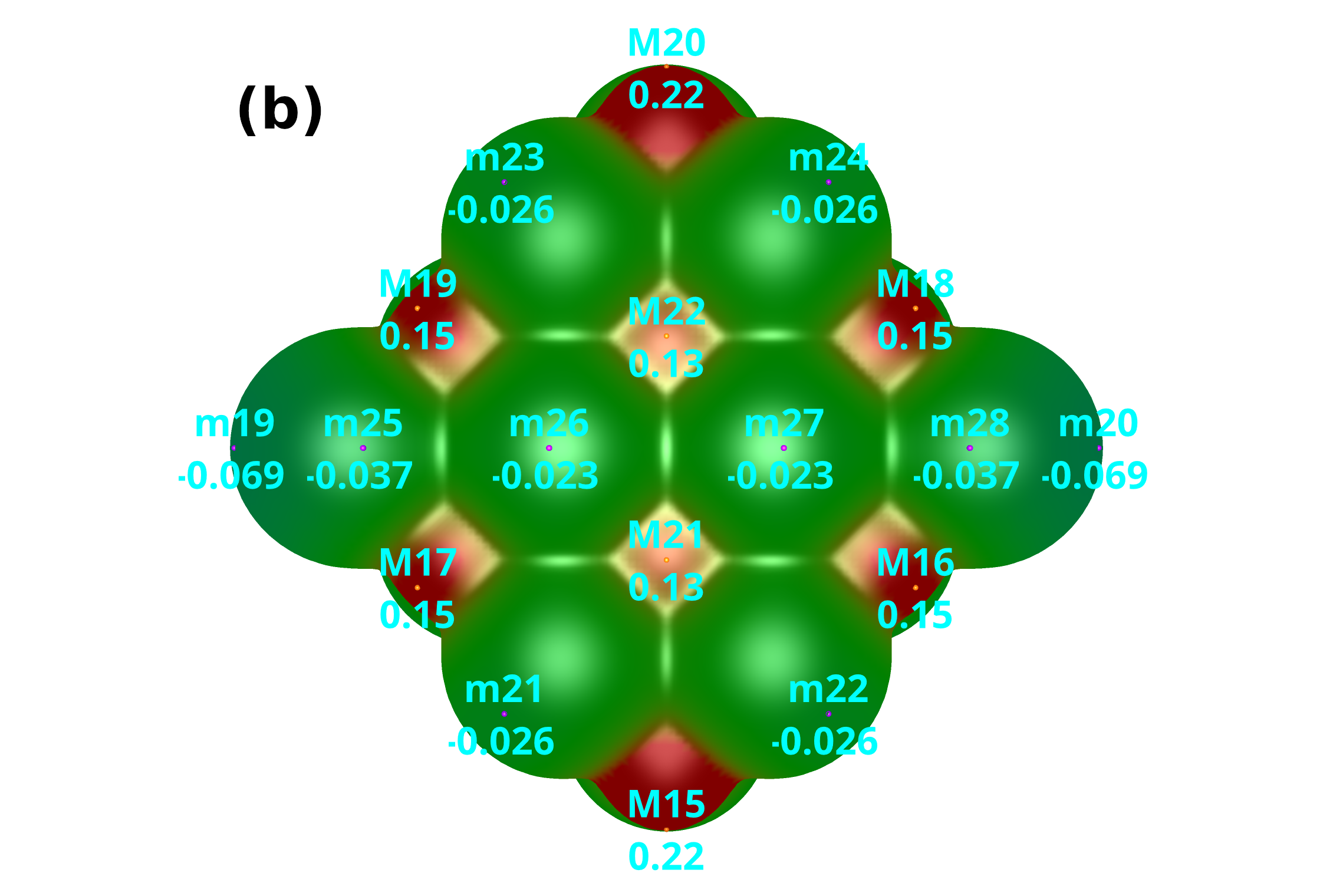}
    
    \vspace*{3cm}
    
    \hspace*{-1.7cm}\includegraphics[width=0.67\textwidth]{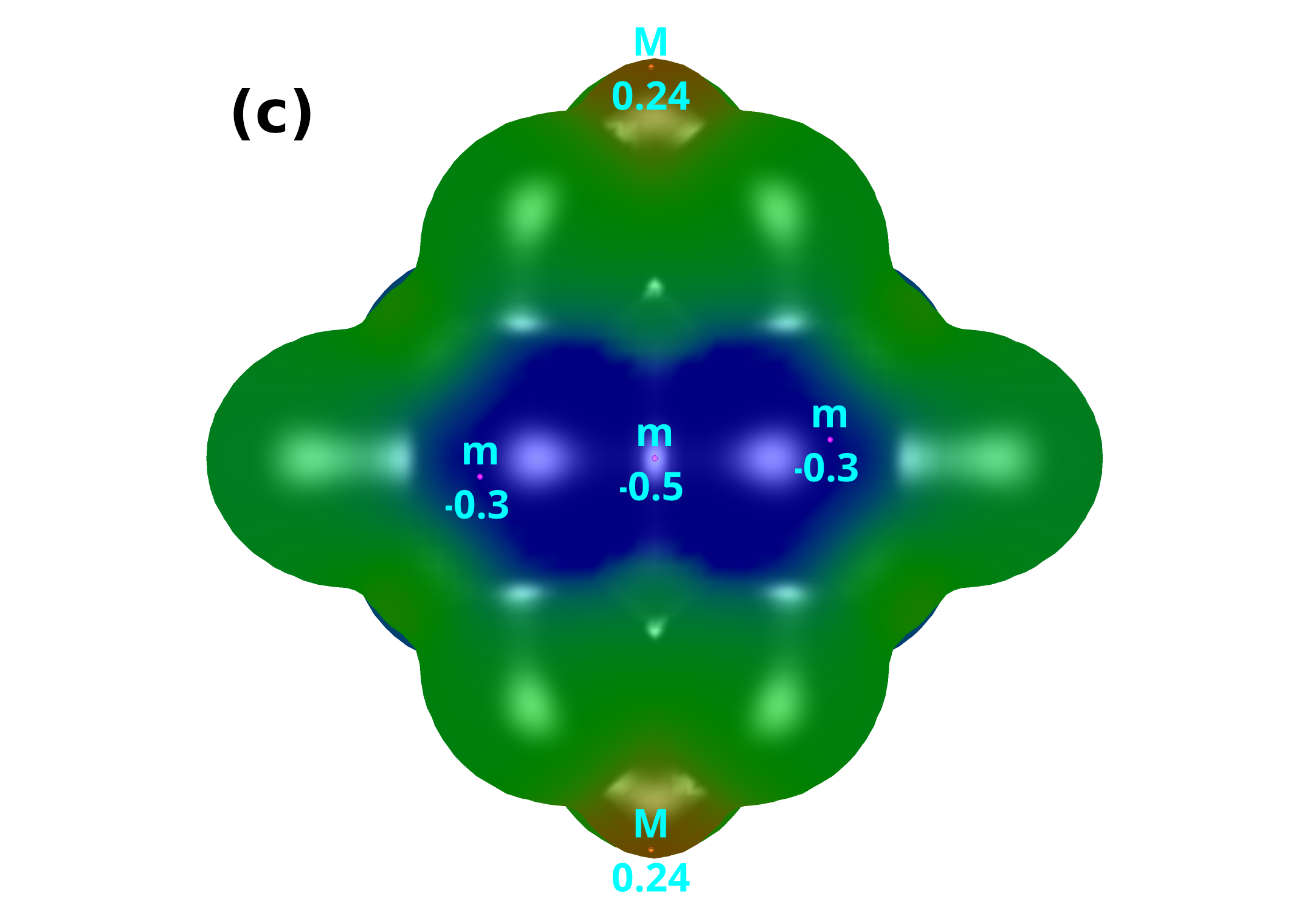}
    \hspace*{-1.5cm}\includegraphics[width=0.67\textwidth]{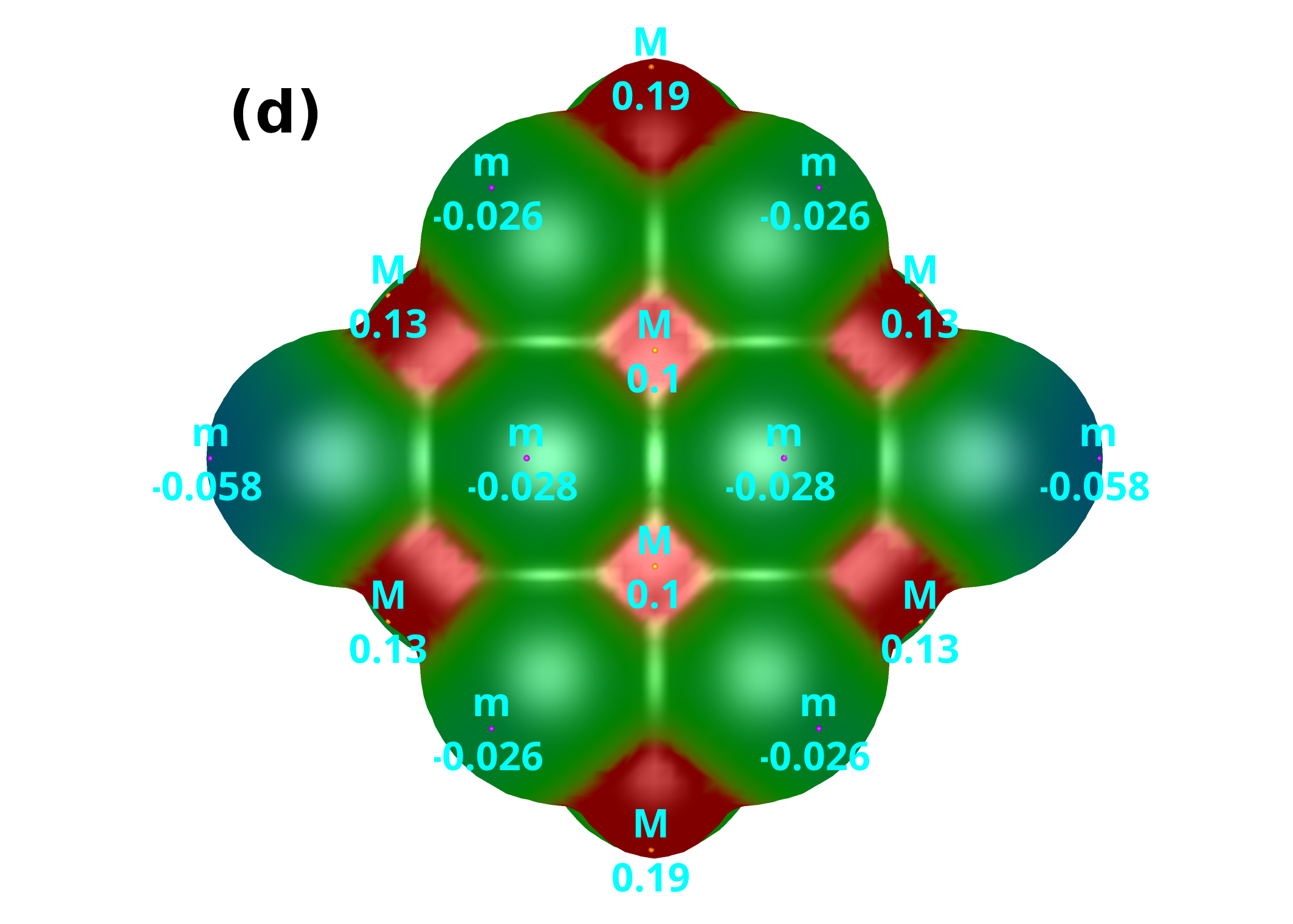}
    \caption{DAM expansion of the molecular electrostatic potential of the cluster (NaCl)$_{24}$ computed at PBE level on the molecular surface. Local maxima and minima shown. 
    (a) \isDZ  $l_{max}= 0$, (b) \isDZ  $l_{max}= 10$, 
    (c)  caDZ  $l_{max}= 0$, (d)  caDZ  $l_{max}= 10$
    \label{fig1} }
\end{figure}

The convergence of the expansion in Eq.~(\ref{eq:1.4}) is illustrated in Figure~\ref{fig1} for the \isDZ and Dunning caDZ basis sets.\cite{Woon1993a,Prascher2011} The figure compares the MESP mapped on the molecular surface, defined by the 0.001 a.u. electron-density isosurface, for two expansions with $l_{\max}=0$ and $l_{\max}=10$. The upper panels correspond to \isDZ, whereas the lower panels show the results obtained with Dunning caDZ. The local extrema on the molecular surface are also indicated together with the corresponding MESP values. The rapid convergence achieved with the \isDZ basis set supports the description of these systems as assemblies of nearly spherical ions. By contrast, the conventional Dunning caDZ basis set exhibits a much slower convergence.

Table~\ref{tab4} presents results for several NaCl clusters, focusing on three key properties: the band gap, lattice energy, and interionic distance. As the cluster size increases, all three properties converge systematically toward the corresponding experimental values. The computational times, measured on a system equipped with an Intel Xeon Platinum 8362 CPU running at 2.80~GHz, are also reported.

Having established the reliability of the \isXZ/\isAJ combination at the RI-PBE level for NaCl clusters, the study was extended to other (MX)$_n$ clusters; and $n = 24$, 90, and 224. Table~\ref{tab5} reports the calculated band gaps, lattice energies, and interionic distances for clusters of these systems with $n = 24$. Table~\ref{tab6} presents the corresponding results for clusters with $n = 90$ together with the Mulliken and L\"owdin atomic charges. The calculated ionic charges remain close to the expected $\pm1$ values, except L\"owdin charges of LiH and KH, as illustrated in Table~\ref{tab6}. Furthermore, the band gaps, lattice energies, and interionic distances computed with the \isXZ basis sets systematically approach the experimental values as both the basis-set cardinality and the cluster size increase.\cite{Alherz2025}

Finally, it is worth mentioning that, in all the systems investigated using the \isXZ basis sets, the frontier molecular orbitals, both HOMO and LUMO, exhibit predominantly anionic character and are largely localized on the X$^-$ ions. A detailed analysis of this feature is beyond the scope of the present work and will be reported elsewhere.

\bibliography{references.bib}

\newpage


\begin{table}[H]
\caption{Composition of Dunning and SIGMA basis sets for ionic atoms. Number of exponents, primitive and contracted functions. \label{tab1}}
\begin{center}
\begin{tabular}{|l|rl|rl|}
\hline
\hline
BS  & $\#$ & Primitives & $\#$  & Contractions \\
\hline
 & \multicolumn{4}{c|}{H$^-$(aXZ) and Li$^+$(cXZ)} \\
\hline
    aDZ         &  11 & ( 5s,  2p)                 &  9 & [ 3s,  2p ]             \\
    cDZ         &  30 & (10s,  5p, 1d)             & 18 & [ 4s,  3p, 1d ]         \\
    aTZ         &  25 & ( 6s,  3p, 2d)             & 23 & [ 4s,  3p, 2d ]         \\
    cTZ         &  61 & (13s,  7p, 3d, 1f)         & 43 & [ 6s,  5p, 3d, 1f ]     \\
    aQZ         &  48 & ( 7s,  4p, 3d, 2f)         & 46 & [ 5s,  4p, 3d, 2f ]     \\
    cQZ         &  90 & (15s,  9p, 5d, 3f, 1g)     & 84 & [ 8s,  7p, 5d, 3f, 1g ] \\
\hline
i$\sigma$DZ1    &  13 & ( 7s,  2p)                 &  5 & [ 2s,  1p ]             \\
i$\sigma$TZ1    &  27 & ( 8s,  3p, 2d)             & 14 & [ 3s,  2p, 1d ]         \\
i$\sigma$QZ1    &  50 & ( 9s,  4p, 3d, 2f)         & 30 & [ 4s,  3p, 2d, 1f ]     \\
\hline
 & \multicolumn{4}{c|}{F$^-$(aXZ) and Na$^+$(cXZ) } \\
\hline
    aDZ         &  35 & (10s,  5p, 2d)             &  23  &[ 4s,  3p, 2d ]            \\
    cDZ         &  49 & (13s,  9p, 2d)             &  27  &[ 5s,  4p, 2d ]            \\
    aTZ         &  58 & (11s,  6p, 3d, 2f)         &  46  &[ 5s,  4p, 3d, 2f ]        \\
    cTZ         &  83 & (18s, 10p, 4d, 2f)         &  59  &[ 7s,  6p, 4d, 2f ]        \\
    aQZ         &  93 & (13s,  7p, 4d, 3f, 2g)     &  80  &[ 6s,  5p, 4d, 3f, 2g ]    \\
    cQZ         & 143 & (22s, 15p, 6d, 4f, 2g)     & 109  &[ 9s,  8p, 6d, 4f, 2g ]     \\
\hline
i$\sigma$DZ1    &  51 & (11s,  7p, 2d)             &  14  &[ 3s, 2p, 1d ]             \\
i$\sigma$TZ1    &  65 & (12s,  8p, 3d, 2f)         &  30  &[ 4s, 3p, 2d, 1f ]         \\
i$\sigma$QZ1    &  90 & (14s,  9p, 4d, 3f, 2g)     &  55  &[ 5s, 4p, 3d, 2f, 1g ]     \\
\hline
 & \multicolumn{4}{c|}{Cl$^-$(aXZ) and K$^+$(pwcXZ)} \\
\hline
    aDZ         &  50 & (13s,  9p,  2d)            &  27  &[ 5s,  4p, 2d ]          \\
  pwcDZ         &  72 & (16s, 12p,  6d)            &  36  &[ 6s,  5p, 3d ]          \\
    aTZ         &  75 & (16s, 10p,  3d, 2f)        &  50  &[ 6s,  5p, 3d, 2f ]      \\
  pwcTZ         & 133 & (25s, 18p,  8d, 2f)        &  68  &[ 8s,  7p, 5d, 2f ]      \\
    aQZ         & 112 & (17s, 12p,  4d, 3f, 2g)    &  84  &[ 7s,  6p, 4d, 3f, 2g ]  \\
\hline
i$\sigma$DZ1    &  57 & (14s, 11p, 2d)             &  18    &[ 4s, 3p, 1d ]             \\
i$\sigma$TZ1    &  81 & (17s, 12p, 3d, 2f)         &  34    &[ 5s, 4p, 2d, 1f ]         \\
i$\sigma$QZ1    & 120 & (18s, 14p, 4d, 3f, 2g)     &  59    &[ 6s, 5p, 3d, 2f, 1g ]     \\
\hline
\end{tabular}
\end{center}
\end{table}



\begin{table}[H]
\setlength{\tabcolsep}{12pt}
\caption{
\isXZ ground state energies (E$_h$) of ions at the HF, CISD. Comparison with Dunning XZ
(aug-VXZ anions and core-VXZ cation, pwvCXZ\cite{Hill2017} for K$^+$). $\Delta = E(\mbox{a(c)XZ})-E(\mbox{\isXZ})$; positive values indicates better sigma performance, differences in mE$_h$
\label{tab2} }
\begin{center}
\begin{tabular}{|l|cccc|}
\hline
BS($\#$c sig/Dun)  & HF & $\Delta$ HF & CISD & $\Delta$ CISD  \\
\hline    
\multicolumn{1}{|c|}{H$^-$}  &  Num$^a$=-0.48793  & \multicolumn{3}{c|}{} \\ 
\hline     
i$\sigma$DZ1(5/9)     & -0.48786  &  1.08 &   -0.52410 &  0.08 \\
i$\sigma$TZ1(14/23)   & -0.48791  &  0.27 &   -0.52667 &  0.10 \\
i$\sigma$QZ1(30/46)   & -0.48792  &  0.11 &   -0.52738 &  0.24 \\
\hline
\multicolumn{1}{|c|}{Li$^+$}  &  Num$^a$=-7.23642  & \multicolumn{3}{c|}{} \\ 
\hline    
i$\sigma$DZ1(5/18)    & -7.23607  & -0.06 &   -7.27204 &  2.83 \\
i$\sigma$TZ1(14/43)   & -7.23630  & -0.08 &   -7.27729 &  0.39 \\
i$\sigma$QZ1(30/84)   & -7.23637  & -0.01 &   -7.27875 &  0.04 \\
\hline
\multicolumn{1}{|c|}{F$^-$}  &  Num$^a$=-99.45945  & \multicolumn{3}{c|}{} \\ 
\hline    
i$\sigma$DZ1(14/23)   & -99.45747 & 29.19 &  -99.67458 & 21.23 \\
i$\sigma$TZ1(30/46)   & -99.45873 &  7.93 &  -99.73265 &  4.80 \\
i$\sigma$QZ1(55/80)   & -99.45919 &  1.72 &  -99.75390 &  0.21 \\
\hline  
\multicolumn{1}{|c|}{Na$^+$}  &  Num$^a$=-161.67696  & \multicolumn{3}{c|}{} \\ 
\hline    
i$\sigma$DZ1(14/27)   &-161.67508 &  4.01 & -161.88365 &  26.89 \\
i$\sigma$TZ1(30/59)   &-161.67631 &  0.29 & -161.94489 &   6.43 \\
i$\sigma$QZ1(55/109)  &-161.67679 &  0.07 & -161.96814 &   1.59 \\
\hline
\multicolumn{1}{|c|}{Cl$^-$}  &  Num$^a$=-459.57693 & \multicolumn{3}{c|}{} \\ 
\hline
i$\sigma$DZ1(18/27)   &-459.57493 & 11.27 & -459.73277 &   6.43 \\
i$\sigma$TZ1(34/50)   &-459.57647 &  2.99 & -459.78545 &   1.12 \\
i$\sigma$QZ1(59/84)   &-459.57678 &  0.43 & -459.80432 &   0.26 \\
\hline
\multicolumn{1}{|c|}{K$^+$}  &  Num==-599.01758& \multicolumn{3}{c|}{} \\ 
\hline
i$\sigma$DZ(18/36)    &-599.01596 &  4.10 & -599.18730 &74.50 \\
i$\sigma$TZ(34/68)    &-599.01727 & -0.41 & -599.24703 &32.38 \\
i$\sigma$QZ(59/118)   &-599.01744 & -0.36 & -599.27004 &13.60 \\
\hline
\end{tabular}    
\end{center}
$^a$ Numerical result from Ref\cite{Koga1995}
\end{table}

\newpage
\begin{table}[H]
    \setlength{\tabcolsep}{8pt}
    \caption{
        Properties of the (NaCl)$_{24}$ (4$\times$4$\times$3) cluster: gap (eV), ionic lattice energies (kJ/mol), and bond distance (\AA) using \isXZ (X=D, T and Q), and \isAJ and \isAMP
            at the RI-HF, RI-PBE and RI-MP2 levels. Computation time for a single-point calculation of the optimized geometry in hours.\label{tab3}}
    \begin{center}
        \begin{tabular}{|l|c|c|c|c|c|}
            \hline
            BS($\#$p / $\#$c)  & $\#$ABS J/MP2 & gap & Latt.E & d & Time \\
            \hline 
            \multicolumn{6}{|c|}{RI-HF } \\ 
            \hline   
            \isDZ(2208/768)  &6792/---& 23.0 & 701.4 & 2.818 & 0.83 \\
            i$\sigma$TZ1(3360/1536) &6792/---& 18.0 & 705.8 & 2.803 & 4.30 \\
            i$\sigma$QZ1(5040/2736) &6792/---& 16.8 & 708.4 & 2.796 & 24.6 \\
            \hline    
            \multicolumn{6}{|c|}{RI-PBE} \\ 
            \hline   
            \isDZ(2208/768)  &6792/---& 12.3 & 725.7 & 2.764 & 0.10 \\
            i$\sigma$TZ1(3360/1536) &6792/---&  8.6 & 728.4 & 2.759 & 0.27 \\
            i$\sigma$QZ1(5040/2736) &6792/---&  7.7 & 733.7 & 2.752 & 0.82 \\
            \hline    
            \multicolumn{6}{|c|}{RI-MP2} \\ 
            \hline   
            \isDZ(2208/768)  &6792/13656&  --- & 707.5 & 2.794 & 1.52 \\
            i$\sigma$TZ1(3360/1536) &6792/13656&  --- & 722.9 & 2.753 & 8.53 \\
            i$\sigma$QZ1(5040/2736) &6792/13656&  --- & 731.8 & 2.745 & 41.3 \\
            \hline    
        \end{tabular}
    \end{center}
    Exp value: gap = 8.6 eV\cite{Roessler1967}, Latt. E = 787 kJ/mol\cite{NIST_NaCl_2026}, $d = 2.820$ \AA\cite{Nickels1949}
\end{table}

\begin{table}[H]
\setlength{\tabcolsep}{8pt}
\caption{
    Properties of the (NaCl)$_n$ clusters , gap (eV), ionic bond energies (kJ/mol) and bond distance (\AA)~using i$\sigma$XZ1 (X=D, T and Q), and i$\sigma$auxJ ABS at the RI-PBE level of calculation. Computation time for a single-point calculation in hours.\label{tab4}
}
\begin{center}
    \begin{tabular}{|l|c|c|c|c|c|}
        \hline
        BS($\#$p/$\#$c)  & $\#$ABS & gap & Latt. E & $d$ & Time \\
        \hline 
        \multicolumn{6}{|c|}{(NaCl)$_{2}$ (2$\times$2$\times$1)} \\ 
        \hline   
        \isDZ(184/64)  &566& 16.0 & 630.3 & 2.552 & - \\
        i$\sigma$TZ1(280/128) &566& 12.0 & 638.1 & 2.549 & - \\
        i$\sigma$QZ1(420/228) &566& 10.3 & 646.1 & 2.543 & - \\
        \hline    
        \multicolumn{6}{|c|}{(NaCl)$_{4}$ (2$\times$2$\times$2)} \\ 
        \hline   
        \isDZ(368/128)  &1132& 14.5 & 679.5 & 2.667 & - \\
        i$\sigma$TZ1(560/256)  &1132& 10.9 & 685.3 & 2.663 & - \\
        i$\sigma$QZ1(840/456)  &1132&  9.9 & 692.2 & 2.657 & - \\
        \hline    
        \multicolumn{6}{|c|}{(NaCl)$_{8}$ (4$\times$4$\times$1)} \\ 
        \hline   
        \isDZ(736/256)  &2264& 13.7 & 687.6 & 2.657 & - \\
        i$\sigma$TZ1(1120/512) &2264&  9.5 & 690.8 & 2.653 & - \\
        i$\sigma$QZ1(1680/912) &2264&  8.4 & 697.7 & 2.643 & - \\
        \hline    
        \multicolumn{6}{|c|}{(NaCl)$_{24}$ (4$\times$4$\times$3)} \\ 
        \hline   
        \isDZ(2208/768)  &6792& 12.3 & 725.7 & 2.764 & 0.1 \\
        i$\sigma$TZ1(3360/1536) &6792&  8.6 & 728.4 & 2.759 & 0.3 \\
        i$\sigma$QZ1(5040/2736) &6792&  7.7 & 733.7 & 2.752 & 0.8 \\
        \hline    
        \multicolumn{6}{|c|}{(NaCl)$_{90}$ (6$\times$6$\times$5)} \\ 
        \hline  
        \isDZ(8280/2880)   &25470& 11.8 & 742.0 & 2.800 &  1.1 \\ 
        i$\sigma$TZ1(13140/5760)  &25470&  7.8 & 743.8 & 2.794 &  3.2 \\ 
        i$\sigma$QZ1(16020/10260) &25470&  6.8 & 748.4 & 2.789 &  6.5 \\ 
        \hline    
        \multicolumn{6}{|c|}{(NaCl)$_{224}$ (8$\times$8$\times$7)} \\ 
        \hline  
        \isDZ(20608/7168)  &63392& 11.7 & 748.7 & 2.816 &  6.8  \\ 
        i$\sigma$TZ1(27328/14336) &63392&  7.3 & 748.9 & 2.810 & 27.8  \\ 
        i$\sigma$QZ1(39872/25536) &63392&  6.5 & 754.4 & 2.801 & 44.3  \\
        \hline    
        \multicolumn{6}{|c|}{(NaCl)$_{450}$ (10$\times$10$\times$9)} \\ 
        \hline  
        \isDZ(41400/14400) &127350& 11.7 & 752.3 & 2.822  & 32.8 \\ 
        i$\sigma$TZ1(54900/28800) &127350&  7.3 & 753.7 & 2.819  & 124. \\ 
        i$\sigma$QZ1(94500/51300) &127350&  6.4 & 757.7 & 2.814  & 267. \\ 
        \hline  
        \multicolumn{6}{|c|}{(NaCl)$_{792}$ (12$\times$12$\times$11)} \\ 
        \hline    
        \isDZ(76824/25344) &224136& 11.7 & 754.5 & 2.831  & 121. \\ 
        i$\sigma$TZ1(96624/50688) &224136&  7.3 & 755.8 & 2.828  & 379. \\ 
        \hline    
    \end{tabular}
\end{center}
Exp value: gap = 8.6 eV, Latt. E = 787 kJ/mol, $d = 2.820$ \AA
\end{table}


\begin{table}[H]
\setlength{\tabcolsep}{8pt}
\caption{
        Properties of the (MX)$_{24}$ clusters (4$\times$4$\times$3), gap (eV), ionic bond energies (kJ/mol) and bond distance (\AA)~using \isXZ (X=D, T and Q), and \isAJ  ABS at the RI-PBE level of calculation.\label{tab5}
    }
\begin{center}
\begin{tabular}{|l|c|c|c|c|c|}
\hline
MX &  BS($\#$p/$\#$c)  & $\#$ABS & gap & Latt. E & $d$ \\
\hline    
LiH  & i$\sigma$DZ1(624/240)   &3552& 5.0 & 854.8 & 1.961 \\
LiH  & i$\sigma$TZ1(1296/672)  &3552& 3.5 & 873.7 & 1.909 \\
LiH  & i$\sigma$QZ1(2400/1440) &3552& 3.4 & 882.0 & 1.904 \\
\hline    
NaH  & i$\sigma$DZ1(1536/456)   &4608& 5.9 & 771.2 & 2.313 \\
NaH  & i$\sigma$TZ1(2208/1056)  &4608& 3.1 & 775.7 & 2.301 \\
NaH  & i$\sigma$QZ1(3360/2040)  &4608& 3.1 & 778.2 & 2.308 \\
\hline   
KH  & i$\sigma$DZ1(1680/552)    &6264& 5.8 & 665.0 & 2.770 \\
KH  & i$\sigma$TZ1(2592/1152)   &6264& 2.7 & 666.4 & 2.745 \\
KH  & i$\sigma$QZ1(4080/2136)   &6264& 2.6 & 672.8 & 2.739 \\
\hline   
LiF  & i$\sigma$DZ1(1536/456)   &4536& 22.1 & 974.0 & 1.973 \\
LiF  & i$\sigma$TZ1(2208/1056)  &4536& 14.6 & 977.9 & 1.957 \\
LiF  & i$\sigma$QZ1(3360/2040)  &4536& 11.5 & 983.1 & 1.947 \\
\hline    
NaF  & i$\sigma$DZ1(2448/672)   &5592& 23.1 & 874.5 & 2.266 \\
NaF  & i$\sigma$TZ1(3120/1440)  &5592& 14.1 & 872.6 & 2.269 \\
NaF  & i$\sigma$QZ1(4320/2640)  &5592&  9.6 & 873.3 & 2.268 \\
\hline    
KF  & i$\sigma$DZ1(2592/768)    &7248& 15.1 & 766.5 & 2.625 \\
KF  & i$\sigma$TZ1(3504/1536)   &7248& 13.6 & 765.6 & 2.627 \\
KF  & i$\sigma$QZ1(5040/2736)   &7248& 10.0 & 777.1 & 2.617 \\
\hline    
LiCl  & i$\sigma$DZ1(1680/552)  &5736& 12.1 & 779.5 & 2.522 \\
LiCl  & i$\sigma$TZ1(2592/1152) &5736&  9.2 & 788.3 & 2.500 \\
LiCl  & i$\sigma$QZ1(4080/2136) &5736&  8.2 & 797.9 & 2.476 \\
\hline    
KCl  & i$\sigma$DZ1(2736/864)   &8448& 13.0 & 648.3 & 3.124 \\
KCl  & i$\sigma$TZ1(3888/1632)  &8448&  8.9 & 649.3 & 3.119 \\
KCl  & i$\sigma$QZ1(5760/2832)  &8448&  7.5 & 657.7 & 3.098 \\
\hline    
\end{tabular}
\end{center}
\end{table}



\begin{table}[H]
\setlength{\tabcolsep}{8pt}
 \caption{
            Properties of the (MX)$_{90}$ clusters (6$\times$6$\times$5), gap (eV), ionic bond energies (kJ/mol), bond distance (\AA)~and Mulliken and L\"owdin charges using i$\sigma$XZ1 (X=D, T and Q), and \isAJ  ABS at the RI-PBE level of calculation. Charges dispersion in parentheses.\label{tab6}
        }
\begin{center}
\begin{tabular}{|l|c|c|c|c|c|c|c|}
\hline
MX &  BS($\#$p/$\#$c)  & $\#$ABS & gap & Latt. E & $d$ & Q$_M$ & Q$_L$\\

\hline   
LiH  & i$\sigma$DZ1(2340/900)   &13320& 4.4 & 875.4 & 1.993 & 1.011~(0.046) & 1.043~(0.051) \\
LiH  & i$\sigma$TZ1(4860/2520)  &13320& 3.5 & 889.2 & 1.946 & 1.011~(0.042) & 1.128~(0.096) \\
LiH  & i$\sigma$QZ1(9000/5400)  &13320& 3.4 & 897.6 & 1.941 & 1.012~(0.193) & 1.247~(0.154) \\
\hline    
NaH  & i$\sigma$DZ1(5760/1710)   &17280& 5.6 & 788.2 & 2.354 & 0.959~(0.060) & 1.007~(0.004) \\
NaH  & i$\sigma$TZ1(8280/3960)   &17280& 3.1 & 789.9 & 2.344 & 1.032~(0.053) & 1.007~(0.004) \\
NaH  & i$\sigma$QZ1(12600/7650)  &17280& 3.1 & 792.3 & 2.349 & 1.025~(0.178) & 1.007~(0.004) \\
\hline   
KH  & i$\sigma$DZ1(6300/2070)   &23490& 5.8 & 679.5 & 2.812 & 1.010~(0.012) & 1.070~(0.022) \\
KH  & i$\sigma$TZ1(9720/4320)   &23490& 2.7 & 679.3 & 2.791 & 1.008~(0.034) & 1.241~(0.094) \\
KH  & i$\sigma$QZ1(15300/8010)  &23490& 2.6 & 685.7 & 2.783 & 1.020~(0.046) & 1.482~(0.132) \\
\hline   
LiF  & i$\sigma$DZ1(5760/1710)   &17010& 21.6 & 992.6 & 2.008 & 1.005~(0.003) & 1.019~(0.004) \\
LiF  & i$\sigma$TZ1(8280/7650)   &17010& 13.4 & 996.1 & 1.990 & 1.004~(0.004) & 1.040~(0.010)  \\
LiF  & i$\sigma$QZ1(12600/39600) &17010& 10.2 &1000.9 & 1.981 & 1.003~(0.006) & 1.082~(0.021)  \\
\hline    
NaF  & i$\sigma$DZ1(9180/2520)   &20970& 22.5 & 893.6 & 2.297 & 1.005~(0.002) & 1.019~(0.004) \\
NaF  & i$\sigma$TZ1(11700/5400)  &20970& 13.5 & 891.3 & 2.300 & 1.006~(0.004) & 1.039~(0.009) \\
NaF  & i$\sigma$QZ1(16200/9900)  &20970&  8.8 & 891.6 & 2.298 & 1.007~(0.005) & 1.081~(0.018) \\
\hline    
NaCl & i$\sigma$DZ1(8280/2880)   &25470& 11.8 & 742.0 & 2.800 & 1.007~(0.004) & 1.021~(0.006)  \\ 
NaCl & i$\sigma$TZ1(13140/5760)  &25470&  7.8 & 743.8 & 2.794 & 1.009~(0.006) & 1.050~(0.014)  \\ 
NaCl & i$\sigma$QZ1(16020/10260) &25470&  6.8 & 748.4 & 2.789 & 1.011~(0.007) & 1.132~(0.031)  \\ 
\hline  
KF  & i$\sigma$DZ1(9720/2880)     &27180& 14.8 & 782.7 & 2.662 & 0.975~(0.001) & 0.993~(0.003) \\
KF  & i$\sigma$TZ1(13140/5760)    &27180& 13.4 & 781.5 & 2.663 & 0.977~(0.002) & 1.014~(0.009) \\
KF  & i$\sigma$QZ1(18900/10260)   &27108&  9.5 & 786.5 & 2.657 & 0.967~(0.003) & 1.021~(0.014) \\
\hline    
LiCl  & i$\sigma$DZ1(6300/2070)  &21510& 11.3 & 795.3 & 2.560 & 1.003~(0.006) & 1.094~(0.036) \\
LiCl  & i$\sigma$TZ1(9720/4320)  &21510&  8.4 & 803.0 & 2.534 & 1.005~(0.009) & 1.019~(0.004) \\
LiCl  & i$\sigma$QZ1(15300/8010) &21510&  7.9 & 811.1 & 2.515 & 1.003~(0.009) & 1.037~(0.016) \\
\hline    
KCl  & i$\sigma$DZ1(10260/3240)   &31680& 12.7 & 663.4 & 3.158 & 0.994~(0.002) & 1.020~(0.007) \\
KCl  & i$\sigma$TZ1(14580/6120)   &31680&  8.4 & 663.8 & 3.155 & 0.994~(0.004) & 1.056~(0.014) \\
KCl  & i$\sigma$QZ1(21600/10620)  &31680&  7.1 & 671.6 & 3.133 & 0.989~(0.005) & 1.014~(0.032) \\
\hline    
\end{tabular}
\end{center}
\end{table}



\begin{table}[H]
\setlength{\tabcolsep}{8pt}
\caption{
    Properties of the (MX)$_{224}$ clusters (8$\times$8$\times$7), gap (eV), ionic bond energies (kJ/mol) and bond distance (\AA)~using \isXZ (X=D, T and Q), and \isAJ  ABS at the RI-PBE level of calculation.\label{tab7}
    }
\begin{center}
\begin{tabular}{|l|c|c|c|c|c|}
\hline
MX &  BS($\#$p/$\#$c)  & $\#$ABS & gap & Latt. E & $d$ \\
\hline   
LiH  & i$\sigma$DZ1(5824/2240)   &33152& 4.1 & 884.4 & 2.008 \\
LiH  & i$\sigma$TZ1(11648/6272)  &33152& 3.0 & 895.5 & 1.963 \\
LiH  & i$\sigma$QZ1(22400/13440) &33152& 2.8 & 903.8 & 1.957 \\
\hline    
NaH  & i$\sigma$DZ1(14336/4256)   &43008& 4.0 & 796.8 & 2.367 \\
NaH  & i$\sigma$TZ1(20608/9856)   &43008& 3.1 & 795.7 & 2.364 \\
NaH  & i$\sigma$QZ1(31360/19040)  &43008& 3.1 & 797.9 & 2.356 \\
\hline   
KH  & i$\sigma$DZ1(15680/5152)   &42336& 5.7 & 685.6 & 2.830  \\
KH  & i$\sigma$TZ1(24192/10752)  &42336& 2.6 & 684.5 & 2.811  \\
KH  & i$\sigma$QZ1(38080/19936)  &42336& 2.6 & 690.9 & 2.803  \\
\hline   
LiF  & i$\sigma$DZ1(14336/4256)   &17010& 21.4 & 999.8 & 2.023  \\
LiF  & i$\sigma$TZ1(20608/19040)  &17010& 13.1 &1003.4 & 2.005  \\
LiF  & i$\sigma$QZ1(31360/98560)  &17010&  9.6 &1008.0 & 1.998  \\
\hline    
NaF  & i$\sigma$DZ1(22848/6270)   &52192& 22.3 & 901.2 & 2.318   \\
NaF  & i$\sigma$TZ1(29120/13440)  &52192& 13.4 & 898.8 & 2.314   \\
NaF  & i$\sigma$QZ1(40320/24640)  &52192&  8.5 & 898.9 & 2.313   \\ %
\hline    
KF  & i$\sigma$DZ1(24192/7168)    &67648& 14.7 & 789.3 & 2.678   \\
KF  & i$\sigma$TZ1(32704/14336)   &67648& 13.3 & 788.0 & 2.679   \\
KF  & i$\sigma$QZ1(47040/25536)   &67648&  9.4 & 792.8 & 2.675   \\
\hline    
LiCl  & i$\sigma$DZ1(15680/5152)  &53536& 11.1 & 801.8 & 2.576  \\
LiCl  & i$\sigma$TZ1(24192/10752) &53536&  8.0 & 809.1 & 2.551  \\
LiCl  & i$\sigma$QZ1(38080/19936) &53536&  7.5 & 816.5 & 2.540  \\
\hline    
KCl  & i$\sigma$DZ1(25536/8064)   &78848& 12.6 & 669.6 & 3.175  \\
KCl  & i$\sigma$TZ1(36288/15232)  &78848&  8.3 & 669.8 & 3.171  \\
KCl  & i$\sigma$QZ1(53760/26432)  &78848&  6.9 & 677.4 & 3.163  \\
\hline    
\end{tabular}
\end{center}
\end{table}

\section*{Appendix: Auxiliary basis sets for \NoCaseChange{i$\sigma$XZ1}}
%
%
The auxiliary basis sets (ABS) \isAJ were designed for calculations involving the M$^+$ and X$^-$ ions in different alkaline halides at HF and DFT levels. For this purpose, HF calculations using the \isQZ basis set for the following diatomic molecules at their equilibrium bond distances in the solids: LiH, NaH, KH, LiF, NaF, KF, LiCl, NaCl, and KCl. Subsequently, for each system, the electron density is partitioned between the two atoms using the DAM partition/expansion algorithm.\cite{FernandezRico2004b} In this partitioning scheme, the total molecular electron density, $D^{AB}$, is expressed as the sum of the densities of two atomic fragments, $D^A$ and $D^B$, which are further expanded as products of real solid spherical harmonics, $z_l^m(\rne)$, multiplied by radial factors, $f_{l,m}(r)$:

\baa \label{eq:2.1}
D^{I}(\rne) = \sum_{l=0}^{l_{max}} \sum_{m=-l}^l f^I_{l,m}(r) \; z_l^m(\rne)
\eaa
where $I = A,B$, and the (unnormalized) regular solid harmonics defined in Eq.\ref{eq:1.2}.

For each ion, the radial factors, $f_{l,m}(r)$, from $l=0$ to $l=6$, are
expanded in terms of $N_l$ primitive Gaussian functions, $g(r;\alpha)$:

\baa \label{eq:2.2}
f^I_{l,m}(r) \simeq \sum_{i=1}^{N_l}  c^I_{ilm} \; g(r;\alpha^I_{i,l}) = \sum_{i=1}^{N_l}  c^I_{ilm} \; \exp^{- \alpha^I_{i,l} r^2} 
\eaa
The expansion coefficients $c^I_{ilm}$ and exponents $\alpha^I_{i,l}$ are optimized by minimizing the self-repulsion energy of atomic fragment $I$:

\baa \label{eq:2.3}
\int d\rne \int d\rne' \; \frac{D^I(\rne) \; D^I(\rne')}{|\rne - \rne'|} & = &
    8 \pi^2 \;
\sum_{l=0}^{\infty} \sum_{m=-l}^l
\frac{1+\delta_{m,0}}{(2l+1)^2} \: \frac{(l+|m|)!}{(l-|m|)!}
\nonumber \\
& \times &
\int_0^\infty dr \; r^{2l} \;
\left[ \, \int_r^\infty dr' \; r' \; f^I_{l,m}(r') \, \right]^2
\eaa
This optimization is carried out by minimizing the functions:

\be \label{eq:2.4}
\Delta^2_{lm} =  \int_0^\infty dt \; t^{2l} \; \left[ \tilde{f}^I_{lm}(t) - \sum_{i=1}^{N_l}  c_{ilm} \;  \frac{ \exp^{\displaystyle{- \alpha^I_{i,l} \,  t^2}} }{2 \alpha_i} \right]^2 
\ee
where:

\be \label{eq:2.5}
\tilde{f}^I_{l,m}(t) = \int_t^\infty dr \; r \; f^I_{l,m}(r) 
\ee

The set of primitive Gaussian functions is optimized to reproduce the individual contributions to the self-repulsion energy, with a relative error of approximately $10^{-10}$ in the integrals of Eq.~(\ref{eq:2.4}). The same tolerance is used to truncate the expansion: only terms with angular momentum $l$ satisfying:

\be \label{eq:2.6}
\int_0^\infty dt \; t^{2l} \; \left[ \tilde{f}^I_{lm}(t) \right]^2 > 10^{-10}
\ee
are retained.
This criterion determines both the number of Gaussian terms, $N_l$, required for each angular momentum $l$, and the highest angular momentum included in the expansion, $l_{\max}$. 

Optimization based on the self-repulsion is preferred over direct optimization of the radial factors themselves because it provides improved accuracy for the Coulomb matrix elements.

Since the optimization is performed simultaneously for the three diatomic molecules associated with a given ion, and the resulting auxiliary functions are shared by all of them, the threshold must be satisfied for every radial factor of that ion appearing in the complete set of diatomics.

The optimized primitive functions are then used directly, without further contraction, to construct the ABS for the ionic clusters. Their composition is summarized in Table~\ref{tabA1}. The \isAJ ABS provide high accuracy and significantly outperform the ORCA\cite{ORCA6} {\it autoAux} procedure, as demonstrated in Tables~\ref{tabA2}--\ref{tabA10} of the Appendix.

In Table~\ref{tabA2}, the performance in HF and PBE calculations is analyzed and found equivalent. According to this, only the HF results are reported in the Tables~\ref{tabA2} to ~\ref{tabA10}.
In all cases, exchange matrix is built without RI procedure.



\newpage

\setcounter{table}{0}
\renewcommand{\thetable}{A\arabic{table}}
\renewcommand{\theHtable}{A\arabic{table}}


\begin{table}[H]
\caption{Composition of auxiliary coulomb SIGMA basis sets \isAJ  for ionic atoms. Number of exponents, primitive and contracted functions. \label{tabA1}}
\begin{center}
\begin{tabular}{|l|rl|}
\hline
\hline
BS  & $\#$ & Primitives \\
\hline
H$^-$      &   96 & ( 8s,  6p,  4d, 3f, 2g, 1h)   \\
Li$^+$     &   52 & ( 9s,  4p,  3d, 1f, 1g)       \\
F$^-$      &  137 & (14s,  9p,  6d, 4f, 3g, 1h)   \\
Na$^+$     &   96 & (14s,  7p,  4d, 3f, 1g, 1h)   \\
Cl$^-$     &  187 & (20s, 11p, 10d, 5f, 3g, 2h)   \\
K$^+$      &  165 & (20s, 11p,  7d, 4f, 3g, 2h)   \\
\hline
\end{tabular}
\end{center}
\end{table}

\newpage

\renewcommand{\baselinestretch}{1.0}

\begin{table}[H]
\setlength{\tabcolsep}{3.5pt}
\vspace*{-10mm}
\begin{center}
\caption{
Comparison of \isAJ and autoAuxN ABS for the (NaCl)$_{24}$  (4$\times$4$\times$3) cluster with the SIGMA basis set \isXZ1 (X=D, T and Q) at the RI-HF and RI-PBE calculation levels. Error in parentheses. \label{tabA2}}
\begin{tabular}{|l|lll|rrr|}
\hline
& \multicolumn{1}{c}{Total Energy} & \multicolumn{1}{c}{E (Na$^+$)}  & \multicolumn{1}{c}{E (Cl$^-$)} & \multicolumn{3}{c|}{$\#$ABS}  \\
\hline
\multicolumn{4}{|c|}{HF 2.82~\AA~isDZ1(2592p/768c)} & Total & Na$^+$ & Cl$^-$ \\
\hline
autoAux 0 &-14916.414765~(3(-3)) & -161.675192~(1(-4)) & -459.575110~(2(-4))&  6720 & 111 & 169\\
autoAux 1 &-14916.411838~(1(-4)) & -161.675078~(3(-6)) & -459.574936~(2(-6))&  7944 & 136 & 195\\
autoAux 2 &-14916.411837~(9(-5)) & -161.675078~(3(-6)) & -459.574936~(2(-6))&  8880 & 152 & 218\\
autoAux 3 &-14916.411799~(6(-5)) & -161.675076~(4(-7)) & -459.574935~(1(-6))& 14880 & 273 & 347\\
\isAJ     &-14916.411745~(3(-6)) & -161.675076~(2(-7)) & -459.574934~(4(-7))&  6792 &  96 & 187\\
reference$^a$  &-14916.411742        & -161.675076        & -459.574934         &   & & \\  
\hline
\multicolumn{4}{|c|}{PBE 2.82~\AA~isDZ1(2592p/768c)} & \multicolumn{3}{c|} {} \\
\hline
autoAux 0 &-14936.173893~(4(-3)) & -161.966869~(1(-4)) & -460.097827~(2(-4))&  6720 & 111 & 169\\
autoAux 1 &-14936.169908~(8(-5)) & -161.966758~(3(-6)) & -460.097658~(2(-6))&  7944 & 136 & 195\\
autoAux 2 &-14936.169910~(8(-5)) & -161.966758~(3(-6)) & -460.097658~(2(-6))&  8880 & 152 & 218\\
autoAux 3 &-14936.169665~(2(-4)) & -161.966755~(3(-7)) & -460.097657~(9(-7))& 14880 & 273 & 347\\
\isAJ     &-14936.169828(2~(-6)) & -161.966755(2~(-7)) & -460.097656(4~(-7))&  6792 &  96 & 187\\
reference$^a$  &-14936.169830        & -161.966755        & -460.097655         &   & & \\  
\hline
\multicolumn{4}{|c|}{HF 2.82~\AA~isTZ1(3504p/1536c)} & \multicolumn{3}{c|} {} \\
\hline
autoAux 0 &-14916.520756~(3(-3)) & -161.676479~(2(-4)) & -459.576652~(2(-4))& 10488 & 176 & 261\\
autoAux 1 &-14916.517551~(1(-4)) & -161.676661~(2(-6)) & -459.576470~(2(-6))& 12240 & 212 & 298\\
autoAux 2 &-14916.517552~(1(-4)) & -161.676479~(2(-6)) & -459.576470~(2(-6))& 13920 & 246 & 334\\
autoAux 3 &-14916.517091~(3(-4)) & -161.676479~(2(-6)) & -459.576468~(2(-8))& 24048 & 442 & 560\\
\isAJ     &-14916.517419~(6(-6)) & -161.676477~(2(-7)) & -459.576468~(2(-7))&  6792 &  96 & 187\\
reference$^a$  &-14916.517413        & -161.676477        & -459.576368         &   & & \\  
\hline
\multicolumn{4}{|c|}{PBE 2.82~\AA~isTZ1(3504p/1536c)} & \multicolumn{3}{c|} {} \\
\hline
autoAux 0 &-14936.436678(4(-3)) & -161.972161(8(-5)) & -460.102561(2(-4))& 10488 & 176 & 261\\
autoAux 1 &-14936.432805(1(-4)) & -161.972082(4(-6)) & -460.102348(2(-6))& 12240 & 212 & 298\\
autoAux 2 &-14936.432801(1(-4)) & -161.972082(4(-6)) & -460.102348(2(-6))& 13920 & 246 & 334\\
autoAux 3 &-14936.432742(7(-5)) & -161.972078(6(-8)) & -460.102346(2(-8))& 24048 & 442 & 560\\
\isAJ     &-14936.432672(4(-6)) & -161.972078(1(-7)) & -460.102346(2(-7))&  6792 &  96 & 187\\
reference$^a$  &-14936.432669        & -161.972078        & -460.102346        &   & & \\  
\hline
\multicolumn{4}{|c|}{HF 2.82~\AA~iQZ1(5040p/2736c)} & \multicolumn{3}{c|} {} \\
\hline
autoAux 0 &-14916.563461~(4(-3)) & -161.676915~(1(-4)) & -459.576977~(2(-4))& 14160 & 237 & 353\\
autoAux 1 &-14916.559955~(1(-4)) & -161.676794~(7(-7)) & -459.576791~(3(-6))& 16296 & 282 & 397\\
autoAux 2 &-14916.559956~(1(-4)) & -161.676794~(7(-7)) & -459.576791~(3(-6))& 19488 & 353 & 459\\
autoAux 3 &-14916.559900~(7(-5)) & -161.676794~(6(-8)) & -459.576785~(2(-7))& 36528 & 661 & 861\\
\isAJ     &-14916.559820~(1(-5)) & -161.676794~(1(-7)) & -459.576786~(1(-7))&  6792 &  96 & 187\\
reference$^a$  &-14916.559830        & -161.676794        & -459.576786         &   & & \\  
\hline
\multicolumn{4}{|c|}{PBE 2.82~\AA~iQZ1(50402p/2736c)} & \multicolumn{3}{c|} {} \\
\hline
autoAux 0 &-14936.530836~(4(-3)) & -161.973025~(1(-4)) & -459.103738~(2(-4))& 14160 & 237 & 353\\
autoAux 1 &-14936.526510~(2(-4)) & -161.972887~(7(-7)) & -459.103529~(5(-6))& 16296 & 282 & 397\\
autoAux 2 &-14936.526520~(2(-4)) & -161.972887~(7(-7)) & -459.103529~(5(-6))& 19488 & 353 & 459\\
autoAux 3 &-14936.527215~(9(-4)) & -161.972886~(9(-9)) & -459.103523~(2(-8))& 36528 & 661 & 861\\
\isAJ     &-14936.526375~(2(-5)) & -161.972886~(1(-9)) & -459.103524~(2(-7))&  6792 &  96 & 187\\
reference$^a$  &-14936.526352        & -161.972886        & -459.103524         &   & & \\  
\hline
\end{tabular}
\end{center}
$^a$ Reference: HF or PBE without RI

\end{table}


\renewcommand{\baselinestretch}{1.2}

\begin{table}[H]
\setlength{\tabcolsep}{3.5pt}
\vspace*{-10mm}
\begin{center}
\caption{
Comparison of \isAJ and autoAuxN ABS for the (LiH)$_{24}$  (4$\times$4$\times$3) cluster with the SIGMA basis set \isXZ1 (X=D, T and Q) at the RI-HF calculation levels. Error in parentheses. \label{tabA3}}
\begin{tabular}{|l|lll|rrr|}
\hline
& \multicolumn{1}{c}{Total Energy} & \multicolumn{1}{c}{E (Li$^+$)} & \multicolumn{1}{c}{E (H$^-$)} & \multicolumn{3}{c|}{$\#$ABS}  \\
\hline 
\multicolumn{4}{|c|}{HF 2.04 ~\AA~isDZ1(624p/240c)} & Total & Li$^+$ & H$^-$ \\ 
\hline 
autoAux 0 &-192.921582~(4(-4)) & -7.236102~(3(-5)) & -0.487877~(1(-5))&  1392 & 38 & 20\\
autoAux 1 &-192.921270~(1(-4)) & -7.236073~(6(-8)) & -0.487863~(4(-8))&  1872 & 48 & 30\\
autoAux 2 &-192.921143~(1(-5)) & -7.236073~(6(-8)) & -0.487863~(4(-8))&  2208 & 48 & 44\\
autoAux 3 &-192.921140~(6(-6)) & -7.236073~(1(-8)) & -0.487863~(1(-8))&  3240 & 71 & 64\\
\isAJ     &-192.921136~(3(-6)) & -7.236073~(3(-8)) & -0.487483~(3(-7))&  3552 & 52 & 96\\
reference$^a$ &-192.921133     & -7.236073         & -0.487863        &   & & \\  
\hline 
\multicolumn{4}{|c|}{HF 2.04 ~\AA~isTZ1(1296p/672c)} & \multicolumn{3}{c|} {} \\ 
\hline 
autoAux 0 &-193.183355~(8(-5)) & -7.236330~(3(-5)) & -0.487905~(2(-8))& 3096 & 75 & 54\\
autoAux 1 &-193.183283~(4(-6)) & -7.236302~(9(-8)) & -0.487905~(2(-8))& 4104 & 93 & 78\\
autoAux 2 &-193.183282~(3(-6)) & -7.236302~(9(-8)) & -0.487905~(2(-8))& 5496 &116 &113\\
autoAux 3 &-193.183277~(3(-6)) & -7.236302~(3(-9)) & -0.487905~(4(-9))& 8064 &178 &158\\
\isAJ     &-193.183281~(1(-6)) & -7.236302~(1(-8)) & -0.487906~(3(-7))& 3552 & 52 & 96\\
reference$^a$ &-193.1832793    & -7.236302        & -0.487905         &   & & \\  
\hline 
\multicolumn{4}{|c|}{HF 2.04 ~\AA~isQZ1(3360p/2040c)} & \multicolumn{3}{c|} {} \\ 
\hline 
autoAux 0 &-193.231330~(6(-5)) & -7.236403~(3(-5)) & -0.487921~(4(-9))& 4420 & 123 & 107\\
autoAux 1 &-193.231272~(4(-6)) & -7.236679~(1(-8)) & -0.487921~(5(-9))& 7608 & 166 & 151\\
autoAux 2 &-193.231269~(2(-6)) & -7.236679~(1(-8)) & -0.487921~(5(-9))& 9528 & 200 & 197\\
autoAux 3 &-193.231274~(7(-6)) & -7.236679~(1(-9)) & -0.487921~(1(-9))&14544 & 313 & 293\\
\isAJ     &-193.231268~(6(-7)) & -7.236679~(1(-8)) & -0.487921~(3(-7))& 3552 &  52 &  96\\
reference$^a$ &-193.231267        & -7.2366794        & -0.487921         &   & & \\  
\hline 
\hline 
\end{tabular}
\end{center}
$^a$ Reference: HF without RI

\end{table}


\begin{table}[H]
\setlength{\tabcolsep}{3.5pt}
\vspace*{-10mm}
\begin{center}
\caption{
Comparison of \isAJ and autoAuxN ABS for the (NaH)$_{24}$  (4$\times$4$\times$3) cluster with the SIGMA basis set \isXZ1 (X=D, T and Q) at the RI-HF calculation levels. Error in parentheses. \label{tabA4}}
\begin{tabular}{|l|lll|rrr|}
\hline
& \multicolumn{1}{c}{Total Energy} & \multicolumn{1}{c}{E (Na$^+$)} & \multicolumn{1}{c}{E (H$^-$)} & \multicolumn{3}{c|}{$\#$ABS}  \\
\hline 
\multicolumn{4}{|c|}{HF 2.49 ~\AA~isDZ1(1536p/456c)} & Total & Na$^+$ & H$^-$ \\ 
\hline 
autoAux 0 &-3898.617425~(1(-3)) & -161.675192~(1(-4)) & -0.487877~(1(-5))&  3144 &111 & 20\\
autoAux 1 &-3898.616582~(1(-4)) & -161.675078~(3(-6)) & -0.487863~(4(-8))&  3984 &136 & 30\\
autoAux 2 &-3898.616522~(7(-5)) & -161.675078~(3(-6)) & -0.487863~(4(-8))&  4704 &152 & 44\\
autoAux 3 &-3898.616527~(7(-5)) & -161.675076~(4(-7)) & -0.487863~(1(-8))&  8088 &273 & 64\\
\isAJ     &-3898.616461~(7(-6)) & -161.675076~(2(-7)) & -0.487863~(3(-7))&  4608 & 96 & 96\\
reference$^a$ &-3898.616454        & -161.675076        & -0.487863         &   & & \\  
\hline 
\multicolumn{4}{|c|}{HF 2.49 ~\AA~isTZ1(2208p/1056c)} & \multicolumn{3}{c|} {}  \\ 
\hline 
autoAux 0 &-3898.786869~(4(-4)) & -161.676388~(8(-5)) & -0.487905~(2(-8))& 5520 &176 & 54\\
autoAux 1 &-3898.786508~(8(-5)) & -161.676316~(4(-6)) & -0.487905~(2(-8))& 6960 &212 & 78\\
autoAux 2 &-3898.786507~(8(-5)) & -161.676316~(4(-6)) & -0.487905~(2(-8))& 8616 &246 &113\\
autoAux 3 &-3898.786467~(4(-5)) & -161.676313~(1(-7)) & -0.487905~(4(-9))&14400 &442 &158\\
\isAJ     &-3898.786423~(1(-6)) & -161.676313~(1(-8)) & -0.487906~(3(-7))& 4608 & 96 & 96\\
reference$^a$ &-3898.786424        & -161.676313        & -0.487905         &   & & \\  
\hline 
\multicolumn{4}{|c|}{HF 2.49 ~\AA~isQZ1(2400p/1440c)} & \multicolumn{3}{c|} {}  \\ 
\hline 
autoAux 0 &-3898.808987~(4(-4)) & -161.676915~(3(-5)) & -0.487921~(4(-9))& 8256 & 237 & 107\\
autoAux 1 &-3898.808330~(8(-5)) & -161.676795~(1(-8)) & -0.487921~(5(-9))&10392 & 282 & 151\\
autoAux 2 &-3898.808326~(2(-6)) & -161.676795~(1(-8)) & -0.487921~(5(-9))&13200 & 353 & 197\\
autoAux 3 &-3898.808326~(7(-6)) & -161.676795~(1(-9)) & -0.487921~(1(-9))&22896 & 661 & 293\\
\isAJ     &-3898.808314~(1(-6)) & -161.676713~(1(-8)) & -0.487921~(3(-7))& 4608 &  96 &  96\\
reference$^a$ &-3898.808314        & -161.676794       & -0.487921         &   & & \\  
\hline 
\hline 
\end{tabular}
\end{center}
$^a$ Reference: HF without RI

\end{table}


\begin{table}[H]
\setlength{\tabcolsep}{3.5pt}
\vspace*{-10mm}
\begin{center}
\caption{
Comparison of \isAJ and autoAuxN ABS for the (KH)$_{24}$  (4$\times$4$\times$3) cluster with the SIGMA basis set \isXZ1 (X=D, T and Q) at the RI-HF calculation levels. Error in parentheses. \label{tabA5}}
\begin{tabular}{|l|lll|rrr|}
\hline
& \multicolumn{1}{c}{Total Energy} & \multicolumn{1}{c}{E (K$^+$)} & \multicolumn{1}{c}{E (H$^-$)} & \multicolumn{3}{c|}{$\#$ABS}  \\
\hline 
\multicolumn{4}{|c|}{HF 2.85 ~\AA~isDZ1(1536p/456c)} & Total & K$^+$ & H$^-$ \\ 
\hline 
autoAux 0 &-14393.863860~(3(-4)) & -599.015966~(6(-6)) & -0.487877~(1(-5))& 4224 &156 & 20\\
autoAux 1 &-14393.863726~(2(-4)) & -599.015964~(4(-6)) & -0.487863~(4(-8))& 4968 &177 & 30\\
autoAux 2 &-14393.863690~(1(-4)) & -599.015964~(4(-6)) & -0.487863~(4(-8))& 5856 &200 & 44\\
autoAux 3 &-14393.863827~(3(-4)) & -599.015960~(2(-7)) & -0.487863~(1(-8))& 9624 &337 & 64\\
\isAJ     &-14393.863584~(2(-5)) & -599.015961~(7(-7)) & -0.487863~(3(-7))& 6264 &165 & 96\\
reference$^a$ &-14393.863564        &  -599.015960       & -0.487863         &   & & \\  
\hline 
\multicolumn{4}{|c|}{HF 2.85 ~\AA~isTZ1(2208p/1056c)} & \multicolumn{3}{c|} {}  \\ 
\hline 
autoAux 0 &-14394.018633~(8(-5)) &  -599.017279~(9(-6)) & -0.487905~(2(-8))& 7032 &239 & 54\\
autoAux 1 &-14394.018705~(2(-4)) &  -599.017276~(6(-6)) & -0.487905~(2(-8))& 8496 &276 & 78\\
autoAux 2 &-14394.018706~(2(-4)) &  -599.017276~(6(-6)) & -0.487905~(2(-8))&10152 &310 &113\\
autoAux 3 &-14394.018667~(1(-4)) &  -599.017270~(7(-8)) & -0.487905~(4(-9))&17280 &562 &158\\
\isAJ    &-14394.0185644~(1(-5)) &  -599.017270~(6(-7)) & -0.487906~(3(-7))& 6264 &165 & 96\\
reference$^a$ &-14394.0185543        &   -599.017270       & -0.487905         &   & & \\  
\hline 
\multicolumn{4}{|c|}{HF 2.85 ~\AA~isQZ1(2400p/1440c)} & \multicolumn{3}{c|} {}  \\ 
\hline 
autoAux 0 &-14394.048431~(8(-5)) & -599.017440~(9(-6)) & -0.487921~(4(-9))&10512 & 331 & 107\\
autoAux 1 &-14394.048502~(2(-4)) & -599.017437~(6(-6)) & -0.487921~(5(-9))&12144 & 355 & 151\\
autoAux 2 &-14394.048498~(1(-4)) & -599.017437~(6(-6)) & -0.487921~(5(-9))&14736 & 417 & 197\\
autoAux 3 &-14394.049177~(8(-4)) & -599.017431~(8(-8)) & -0.487921~(1(-9))&27576 & 856 & 293\\
\isAJ     &-14394.048365~(1(-5)) & -599.017432~(6(-7)) & -0.487921~(3(-7))& 6264 &165 &  96\\
reference$^a$ &-14394.048352        & -599.017431       & -0.487921         &   & & \\  
\hline 
\hline 
\end{tabular}
\end{center}
$^a$ Reference: HF without RI

\end{table}


\begin{table}[H]
\setlength{\tabcolsep}{3.5pt}
\vspace*{-10mm}
\begin{center}
\caption{
Comparison of \isAJ and autoAuxN ABS for the (LiF)$_{24}$  (4$\times$4$\times$3) cluster with the SIGMA basis set \isXZ1 (X=D, T and Q) at the RI-HF calculation levels. Error in parentheses. \label{tabA6}}
\begin{tabular}{|l|lll|rrr|}
\hline
& \multicolumn{1}{c}{Total Energy} & \multicolumn{1}{c}{E (Li$^+$)} & \multicolumn{1}{c}{E (F$^-$)} & \multicolumn{3}{|c|}{$\#$ABS}  \\
\hline 
\multicolumn{4}{|c|}{HF 2.04 ~\AA~isDZ1(1536p/456c)} & Total & Li$^+$ & F$^-$ \\ 
\hline 
autoAux 0 &-2568.431232~(2(-2)) & -7.236102~(3(-5)) & -99.458397~(9(-4))&  3576 & 38 &111\\
autoAux 1 &-2568.410130~(2(-4)) & -7.236073~(6(-8)) & -99.457482~(7(-6))&  4632 & 48 &145\\
autoAux 2 &-2568.410130~(2(-4)) & -7.236073~(6(-8)) & -99.457482~(7(-6))&  5184 & 48 &168\\
autoAux 3 &-2568.409961~(1(-5)) & -7.236073~(1(-8)) & -99.457476~(1(-6))&  8472 & 71 &282\\
\isAJ     &-2568.409985~(1(-5)) & -7.236073~(3(-8)) & -99.457475~(4(-7))&  4536 & 52 &137\\
reference$^a$ &-2568.409974        & -7.236073        & -99.457475  &   & & \\  
\hline 
\multicolumn{4}{|c|}{HF 2.04 ~\AA~isTZ1(2208p/1056c)} & \multicolumn{3}{c|} {}  \\ 
\hline 
autoAux 0 &-2568.469994~(1(-2)) & -7.2363759~(3(-5)) & -99.459314~(6(-4))& 6240 & 75 &185\\
autoAux 1 &-2568.457428~(2(-4)) & -7.2363759~(9(-8)) & -99.458741~(1(-5))& 7776 & 93 &231\\
autoAux 2 &-2568.457429~(2(-4)) & -7.2363759~(9(-8)) & -99.458741~(1(-5))& 9360 &116 &274\\
autoAux 3 &-2568.457160~(2(-5)) & -7.2363759~(3(-9)) & -99.458731~(2(-7))&15312 &178 &460\\
\isAJ     &-2568.457178~(3(-6)) & -7.2363759~(1(-8)) & -99.458731~(7(-8))& 4536 & 52 &137\\
reference$^a$ &-2568.457181        & -7.2363759        & -99.458731       &   & & \\  
\hline 
\multicolumn{4}{|c|}{HF 2.04 ~\AA~isQZ1(2400p/1440c)} & \multicolumn{3}{c|} {}  \\ 
\hline 
autoAux 0 &-2568.499649~(2(-2)) & -7.236403~(3(-5)) & -99.459938~(7(-4))&  4536 & 123 & 255\\
autoAux 1 &-2568.483726~(4(-5)) & -7.236376~(3(-5)) & -99.459194~(2(-6))&  7608 & 166 & 307\\
autoAux 2 &-2568.483722~(3(-5)) & -7.236376~(1(-8)) & -99.459194~(2(-6))&  9528 & 200 & 362\\
autoAux 3 &-2568.483596~(9(-5)) & -7.236376~(1(-9)) & -99.459192~(9(-8))& 14544 & 313 & 711\\
\isAJ     &-2568.483684~(3(-6)) & -7.236376~(1(-8)) & -99.459192~(1(-8))&  3552 &  52 & 137\\
reference$^a$ &-2568.483687        & -7.236376 &  -99.459192        &   & & \\  
\hline 
\hline 
\end{tabular}
\end{center}
$^a$ Reference: HF without RI

\end{table}


\begin{table}[H]
\setlength{\tabcolsep}{3.5pt}
\vspace*{-10mm}
\begin{center}
\caption{
Comparison of \isAJ and autoAuxN ABS for the (NaF)$_{24}$  (4$\times$4$\times$3) cluster with the SIGMA basis set \isXZ1 (X=D, T and Q) at the RI-HF calculation levels. Error in parentheses. \label{tabA7}}
\begin{tabular}{|l|lll|rrr|}
\hline
& \multicolumn{1}{c}{Total Energy} & \multicolumn{1}{c}{E (Na$^+$)} & \multicolumn{1}{c}{E (F$^-$)} & \multicolumn{3}{|c|}{$\#$ABS}  \\
\hline 
\multicolumn{4}{|c|}{HF 2.49 ~\AA~isDZ1(1536p/456c)} & Total & Li$^+$ & F$^-$ \\ 
\hline 
autoAux 0 &-6274.632514~(2(-2)) & -161.675192~(1(-4)) & -99.458397~(9(-4))&  5328 &111 &111\\
autoAux 1 &-6274.617037~(2(-4)) & -161.675078~(3(-6)) & -99.457482~(7(-6))&  6744 &136 &145\\
autoAux 2 &-6274.617031~(2(-4)) & -161.675078~(3(-6)) & -99.457482~(7(-6))&  7680 &152 &168\\
autoAux 3 &-6274.616849~(3(-5)) & -161.675076~(4(-7)) & -99.457476~(1(-6))& 13320 &273 &282\\
\isAJ     &-6274.616812~(2(-6)) & -161.675076~(2(-7)) & -99.457475~(4(-7))&  5592 & 96 &137\\
reference$^a$ &-6274.616814        & -161.675076        & -99.457475       &   & & \\  
\hline 
\multicolumn{4}{|c|}{HF 2.49 ~\AA~isTZ1(2208p/1056c)} & \multicolumn{3}{c|} {}  \\ 
\hline 
autoAux 0 &-6274.770019~(9(-3)) & -161.676388~(8(-5)) & -99.459314~(6(-4))& 8664 &176 &185\\
autoAux 1 &-6274.760928~(3(-4)) & -161.676316~(4(-6)) & -99.458741~(1(-5))&10632 &212 &231\\
autoAux 2 &-6274.760934~(3(-4)) & -161.676316~(4(-6)) & -99.458741~(1(-5))&12480 &246 &274\\
autoAux 3 &-6274.760316~(3(-4)) & -161.676313~(1(-7)) & -99.458731~(2(-7))&21648 &442 &460\\
\isAJ     &-6274.760601~(2(-5)) & -161.676313~(1(-8)) & -99.458731~(7(-8))& 5592 & 96 &137\\
reference$^a$ &-6274.760624        & -161.676313        & -99.458731       &   & & \\  
\hline 
\multicolumn{4}{|c|}{HF 2.49 ~\AA~isQZ1(2400p/1440c)} & \multicolumn{3}{c|} {}  \\ 
\hline 
autoAux 0 &-6274.841362~(1(-2)) & -161.676915~(3(-5)) & -99.459938~(7(-4))&11808 & 237 & 255\\
autoAux 1 &-6274.830869~(6(-7)) & -161.676795~(1(-8)) & -99.459194~(2(-6))&14136 & 282 & 307\\
autoAux 2 &-6274.830886~(9(-6)) & -161.676795~(1(-8)) & -99.459194~(2(-6))&17160 & 353 & 362\\
autoAux 3 &-6274.830293~(6(-4)) & -161.676795~(1(-9)) & -99.459192~(9(-8))&32928 & 661 & 711\\
\isAJ     &-6274.830330~(4(-5)) & -161.676794~(1(-8)) & -99.459192~(1(-8))& 5952 & 96 & 137\\
reference$^a$ &-6274.830869         & -161.676794       &  -99.459192       &   & & \\  
\hline 
\hline 
\end{tabular}
\end{center}
$^a$ Reference: HF without RI

\end{table}


\begin{table}[H]
\setlength{\tabcolsep}{3.5pt}
\vspace*{-10mm}
\begin{center}
\caption{
Comparison of \isAJ and autoAuxN ABS for the (KF)$_{24}$  (4$\times$4$\times$3) cluster with the SIGMA basis set \isXZ1 (X=D, T and Q) at the RI-HF calculation levels. Error in parentheses. \label{tabA8}}
\begin{tabular}{|l|lll|rrr|}
\hline
& \multicolumn{1}{c}{Total Energy} & \multicolumn{1}{c}{E (K$^+$)} & \multicolumn{1}{c}{E (F$^-$)} & \multicolumn{3}{c|}{$\#$ABS}  \\
\hline
\multicolumn{4}{|c|}{HF 2.66 ~\AA~isDZ1(2112p/768c)} & Total & K$^+$ & F$^-$ \\ 
\hline 
autoAux 0 & -16770.198850~(1(-2)) &  -599.015966~(6(-6)) & -99.458397~(9(-4))&  7800 & 156 & 169\\
autoAux 1 & -16770.189044~(3(-4)) &  -599.015964~(4(-6)) & -99.457482~(7(-6))&  8928 & 177 & 195\\
autoAux 2 & -16770.189041~(3(-4)) &  -599.015964~(4(-6)) & -99.457482~(7(-6))& 10032 & 200 & 218\\
autoAux 3 & -16770.188706~(7(-5)) &  -599.015960~(2(-7)) & -99.457476~(1(-6))& 16416 & 337 & 347\\
\isAJ     & -16770.188799~(2(-5)) &  -599.015961~(1(-6)) & -99.457475~(4(-7))&  8448 & 165 & 187\\
reference$^a$ & -16770.188778        &  -599.015960        & -99.457475         &   & & \\  
\hline 
\multicolumn{4}{|c|}{HF 2.66 ~\AA~isTZ1(2928p/1536c)} & \multicolumn{3}{c|} {}  \\ 
\hline 
autoAux 0 & -16770.284056~(7(-3)) &-599.0174388~(8(-5)) & -99.459314~(6(-4))& 12000 & 239 & 261\\
autoAux 1 & -16770.277347~(3(-4)) &-599.0174316~(4(-6)) & -99.458741~(1(-5))& 13776 & 276 & 298\\
autoAux 2 & -16770.277311~(3(-4)) &-599.0174316~(4(-6)) & -99.458741~(1(-5))& 15456 & 310 & 334\\
autoAux 3 & -16770.277080~(9(-4)) &-599.0174313~(1(-7)) & -99.458731~(2(-7))& 26928 & 562 & 560\\
\isAJ     & -16770.277697~(2(-5)) &-599.0174313~(6(-8)) & -99.458731~(7(-8))&  8448 & 165 & 187\\
reference$^a$ & -16770.277699        &-599.0174313        & -99.458731       &   & & \\  
\hline 
\multicolumn{4}{|c|}{HF 2.66 ~\AA~isQZ1(4272p/2736c)} & \multicolumn{3}{c|} {}  \\ 
\hline 
autoAux 0 &-16770.331528~(8(-3)) &-599.017440~(9(-6)) & -99.459938~(7(-4))& 14064 & 331 & 255\\
autoAux 1 &-16770.324150~(1(-4)) &-599.017437~(6(-6)) & -99.459194~(2(-6))& 15888 & 355 & 307\\
autoAux 2 &-16770.324134~(1(-4)) &-599.017437~(6(-6)) & -99.459194~(2(-6))& 18696 & 417 & 362\\
autoAux 3 &-16770.323360~(7(-4)) &-599.017431~(8(-8)) & -99.459192~(9(-8))& 37608 & 856 & 711\\
\isAJ     &-16770.323991~(3(-5)) &-599.017432~(6(-7)) & -99.459192~(1(-8))&  7248 & 165 & 137\\
reference$^a$ &-16770.324020        &-599.017431        & -99.459192         &   & & \\  
\hline 
\hline 
\end{tabular}
\end{center}
$^a$ Reference: HF without RI

\end{table}

\newpage

\begin{table}[H]
\setlength{\tabcolsep}{3.5pt}
\vspace*{-10mm}
\begin{center}
\caption{
Comparison of \isAJ and autoAuxN ABS for the (LiCl)$_{24}$  (4$\times$4$\times$3) cluster with the SIGMA basis set \isXZ1 (X=D, T and Q) at the RI-HF calculation levels. Error in parentheses. \label{tabA9}}
\begin{tabular}{|l|lll|rrr|}
\hline
& \multicolumn{1}{c}{Total Energy} & \multicolumn{1}{c}{E (Li$^+$)} & \multicolumn{1}{c}{E (Cl$^-$)} & \multicolumn{3}{|c|}{$\#$ABS}  \\
\hline 
\multicolumn{4}{|c|}{HF 2.57 ~\AA~isDZ1(1536p/456c)} & Total & Li$^+$ & Cl$^-$ \\ 
\hline 
autoAux 0 &-11210.398502~(3(-3)) & -7.236102~(3(-5)) & -459.575110~(2(-4))&  4968 &  38 & 169\\
autoAux 1 &-11210.395719~(4(-5)) & -7.236073~(6(-8)) & -459.574936~(2(-6))&  5832 &  48 & 195\\
autoAux 2 &-11210.395718~(4(-5)) & -7.236073~(6(-8)) & -459.574936~(2(-6))&  6384 &  48 & 218\\
autoAux 3 &-11210.395685~(3(-6)) & -7.236073~(1(-8)) & -459.574935~(1(-6))& 10032 &  71 & 347\\
\isAJ     &-11210.395685~(4(-6)) & -7.236073~(3(-8)) & -459.574934~(4(-7))&  5736 &  52  & 187\\
reference$^a$ &-11210.395682        & -7.236073        & -459.574934         &   & & \\  
\hline 
\multicolumn{4}{|c|}{HF 2.57 ~\AA~isTZ1(2208p/1056c)}  & \multicolumn{3}{c|} {}   \\ 
\hline 
autoAux 0 &-11210.520695~(3(-3)) &  -7.236330~(3(-5)) & -459.576652~(2(-4))&  8064 &  75 & 261\\
autoAux 1 &-11210.518218~(8(-5)) &  -7.236302~(9(-8)) & -459.576470~(2(-6))&  9384 &  93 & 298\\
autoAux 2 &-11210.518227~(9(-5)) &  -7.236302~(9(-8)) & -459.576470~(2(-6))& 10800 & 116 & 334\\
autoAux 3 &-11210.518179~(4(-5)) &  -7.236302~(3(-9)) & -459.576468~(2(-8))& 17712 & 178 & 560\\
\isAJ     &-11210.518149~(1(-5)) &  -7.236302~(9(-9)) & -459.576468~(2(-7))&  5736 &  52 & 187\\
reference$^a$ &-11210.518134        &  -7.236302        & -459.576368         &   & & \\  
\hline 
\multicolumn{4}{|c|}{HF 2.57 ~\AA~isQZ1(336p/2040c)}  & \multicolumn{3}{c|} {}  \\ 
\hline 
autoAux 0 &-11210.586050~(2(-3)) &  -7.236403~(3(-5)) & -459.576975~(2(-4))& 11424 & 123 & 353\\
autoAux 1 &-11210.583912~(1(-4)) &  -7.236376~(1(-8)) & -459.576788~(5(-6))& 13512 & 166 & 397\\
autoAux 2 &-11210.583916~(1(-4)) &  -7.236376~(1(-8)) & -459.576788~(5(-6))& 15816 & 200 & 459\\
autoAux 3 &-11210.584573~(8(-4)) &  -7.236376~(1(-9)) & -459.576794~(1(-8))& 28176 & 313 & 347\\
\isAJ     &-11210.583794~(4(-6)) &  -7.236376~(6(-9)) & -459.576794~(1(-7))&  5736 &  52 & 187\\
reference$^a$ &-11210.583790        &  -7.236376        & -459.576794         &   & & \\  
\hline 
\end{tabular}
\end{center}
$^a$ Reference: HF without RI

\end{table}

\newpage

\begin{table}[H]
\setlength{\tabcolsep}{3.5pt}
\vspace*{-10mm}
\begin{center}
\caption{
Comparison of \isAJ and autoAuxN ABS for the (KCl)$_{24}$  (4$\times$4$\times$3) cluster with the SIGMA basis set \isXZ1 (X=D, T and Q) at the RI-HF calculation levels. Error in parentheses.\label{tabA10}}
\begin{tabular}{|l|lll|rrr|}
\hline
& \multicolumn{1}{c}{Total Energy} & \multicolumn{1}{c}{E (K$^+$)} & \multicolumn{1}{c|}{E (Cl$^-$)} & \multicolumn{3}{c|}{$\#$ABS}  \\
\hline 
\multicolumn{4}{|c|}{HF 3.14 ~\AA~isDZ1(2496p/864c)} & Total & K$^+$ & Cl$^-$ \\ 
\hline 
autoAux 0 & -25411.863231~(3(-3)) &  -599.015966~(6(-6)) & -459.575110~(2(-4))&  7800 & 156 & 169\\
autoAux 1 & -25411.860717~(2(-4)) &  -599.015964~(4(-6)) & -459.574936~(2(-6))&  8928 & 177 & 195\\
autoAux 2 & -25411.860698~(1(-4)) &  -599.015964~(4(-6)) & -459.574936~(2(-6))& 10032 & 200 & 218\\
autoAux 3 & -25411.860674~(1(-4)) &  -599.015960~(2(-7)) & -459.574935~(1(-6))& 16416 & 337 & 347\\
\isAJ     & -25411.860589~(2(-5)) &  -599.015961~(1(-6)) & -459.574934~(4(-7))&  8448 & 165 & 187\\
reference$^a$ & -25411.860565        &  -599.015960        & -459.574934         &   & & \\  
\hline 
\multicolumn{4}{|c|}{HF 3.14 ~\AA~isTZ1(3888p/1634c)}  & \multicolumn{3}{c|} {}  \\ 
\hline 
autoAux 0 & -25411.961047~(2(-3)) &-599.0174388~(8(-5)) & -459.576652~(2(-4))& 12000 & 239 & 261\\
autoAux 1 & -25411.958750~(2(-4)) &-599.0174316~(4(-6)) & -459.576470~(2(-6))& 13776 & 276 & 298\\
autoAux 2 & -25411.958745~(2(-4)) &-599.0174316~(4(-6)) & -459.576470~(2(-6))& 15456 & 310 & 334\\
autoAux 3 & -25411.958153~(4(-4)) &-599.0174313~(1(-7)) & -459.576468~(2(-8))& 26928 & 562 & 560\\
\isAJ     & -25411.958569~(1(-5)) &-599.0174313~(6(-8)) & -459.576468~(2(-7))&  8448 & 165 & 187\\
reference$^a$ & -25411.958555        &-599.0174313        & -459.576368         &   & & \\  
\hline 
\multicolumn{4}{|c|}{HF 3.14 ~\AA~isQZ1(5760p/2832c)} & \multicolumn{3}{c|} {}  \\ 
\hline 
autoAux 0 &-25412.0063~(2(-3)) &-599.017440~(9(-6)) & -459.576975~(2(-4))& 14160 & 331 & 353\\
autoAux 1 &-25412.0044~(3(-4)) &-599.017437~(6(-6)) & -459.576788~(5(-6))& 16296 & 355 & 397\\
autoAux 2 &-25412.0044~(3(-4)) &-599.017437~(6(-6)) & -459.576788~(5(-6))& 19488 & 417 & 459\\
autoAux 3 &-25412.0019~(2(-3)) &-599.017431~(8(-8)) & -459.576794~(1(-8))& 36528 & 856 & 861\\
\isAJ     &-25412.0041~(7(-6)) &-599.017432~(6(-7)) & -459.576794~(1(-7))&  8448 & 165 & 187\\
reference$^a$ &-25412.004138        &-599.017431        & -459.576794         &   & & \\  
\hline 
\end{tabular}
\end{center}
$^a$ Reference: HF without RI

\end{table}

\end{document}